%% file: main.tex
\pdfoutput=1
\documentclass[sigconf]{acmart} %% or use sigconf for publication

\AtBeginDocument{%
  \providecommand\BibTeX{{%
    \normalfont B\kern-0.5em{\scshape i\kern-0.25em b}\kern-0.8em\TeX}}}

\setcopyright{acmlicensed}
\copyrightyear{2024}
\acmYear{2024}
\acmDOI{XXXXXXX.XXXXXXX}

\acmConference[CHI In2Writing Workshop '24]{Make sure to enter the correct
  conference title from your rights confirmation email}{May 2024}{Hawaii, USA}
\usepackage{graphicx}
\usepackage{subcaption}

\usepackage{microtype}
\usepackage{enumitem}
\usepackage{booktabs}
\usepackage{tabularx}
\usepackage{multirow}
\usepackage[frozencache, cachedir=.]{minted}

\setcopyright{none}
\renewcommand\footnotetextcopyrightpermission[1]{}
\begin{document}
\title{\textit{Human-AI Collaborative Taxonomy Construction}: \\A Case Study in Profession-Specific Writing Assistants}

% \affiliation{%

\author{Minhwa Lee}
\affiliation{%
  \institution{University of Minnesota}
  \city{Twin Cities}
  \state{Minnesota}
  \country{USA}}
\email{lee03533@umn.edu}

\author{Zae Myung Kim}
\affiliation{%
  \institution{University of Minnesota}
  \city{Twin Cities}
  \state{Minnesota}
  \country{USA}}
\email{kim01756@umn.edu}

%% Authorship between Vivek and DK...?
\author{Vivek Khetan}
\affiliation{%
  \institution{Accenture Labs}
  \city{San Francisco}
  \state{California}
  \country{USA}}
\email{vivek.a.khetan@accenture.com}

\author{Dongyeop Kang}
\affiliation{%
  \institution{University of Minnesota}
  \city{Twin Cities}
  \state{Minnesota}
  \country{USA}}
\email{dongyeop@umn.edu}
\renewcommand{\shortauthors}{Lee, et al.}

\begin{abstract}

Large Language Models (LLMs) have assisted humans in several writing tasks, including text revision and story generation. However, their effectiveness in supporting domain-specific writing, particularly in business contexts, is relatively less explored. Our formative study with industry professionals revealed the limitations in current LLMs' understanding of the nuances in such domain-specific writing. To address this gap, we propose an approach of human-AI collaborative taxonomy development to perform as a guideline for domain-specific writing assistants. This method integrates iterative feedback from domain experts and multiple interactions between these experts and LLMs to refine the taxonomy. Through larger-scale experiments, we aim to validate this methodology and thus improve LLM-powered writing assistance, tailoring it to meet the unique requirements of different stakeholder needs.

\end{abstract}









\maketitle
%%%%%%%%%%%%%%%%%%%%%%%%%%%%%%%%%%%%%%%%%%%%%%%%

\begin{figure*}[h]
    \includegraphics[width=\linewidth]{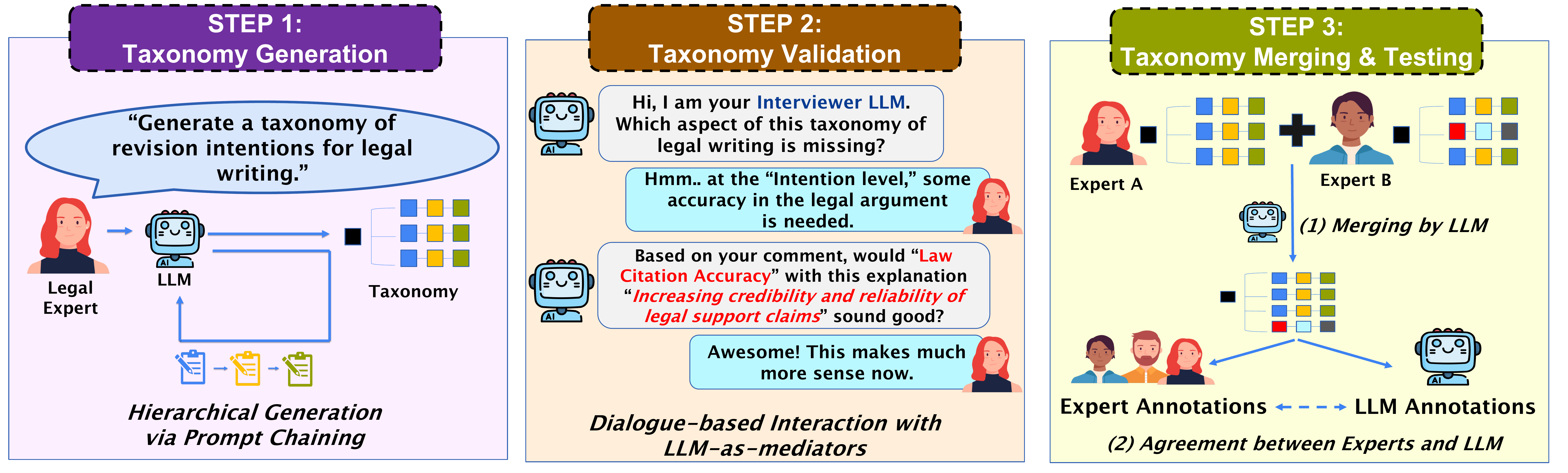}
    \caption{An end-to-end pipeline of our three-step Human-AI collaborative taxonomy construction process. For each step, we portray several design implications for better human-AI interaction strategies that were described in Section 3.}
    \label{fig:pipeline}
\end{figure*}

\input{Sections/Introduction}

\input{Sections/Formative}

\input{Sections/New}

\input{Sections/Results_Conclusion}

\bibliographystyle{ACM-Reference-Format}
\bibliography{ref}

\input{Sections/Appendix}

\end{document}

%% file: Sections/Introduction.tex
\section{Introduction}

% \begin{itemize}
%     \item Recent advancements in modeling techniques lead to human-like capabilities of large language models (LLMs) on many downstream text generation tasks. Those LLMs are now deployed to many AI-driven writing assistants, including but not limited to syntactic and semantic edit revisions (e.g., grammar correction, coherency improvement, etc.) and creative content generation (e.g., story/plot generations).  
%     \item Despite the remarkable performance of the LLM-driven tools, yet understudied in stakeholder-driven writing support with those AI. Can we trust the generated responses of those AI writing assistants when we do not know they can model the social knowledge of the stakeholders?
%     Also, the commercial release of business-specific AI writing assistants developed by the company ‘Writer’. Our work is based on answering the question: what are the essential steps to develop such stakeholder-specific writing assistants that can be provided in an open-source setting?
%     \item Based on our formative study, we found several drawbacks of current LLMs as an assistant tool for business domains that require an in-depth understanding of that domain-specific knowledge (e.g., hidden rule-of-thumb among stakeholders). -> Propose an essential step of "human-AI collaborative approach" of writing guideline generation with rigorous expert validation procedures, for better domain-specific writing assistants. 
% \end{itemize} 

The emergence of large language models (LLMs) has significantly enriched the writing support tools, covering text revisions \cite{du-etal-2022-read, raheja2023coedit, schick2022peer} to creative content generation \cite{Mirowski2022CoWritingSA, chakrabarty2022help, ippolito2022creative}. However, the deployment of LLMs in tailored writing assistance to specific stakeholder needs (e.g., legal writing) remains relatively unexplored. This oversight highlights a concern about the reliability of AI-driven tools in meeting the nuanced demands of domain-specific writing. Also, commercial offerings, such as the company Writer's LLM-powered writing assistants for business sectors (e.g., Palmyra \cite{Palmyra}), are not widely accessible. This raises questions about the essential steps to develop publicly available AI writing assistants that cater to different stakeholders' unique needs and knowledge. 

Motivated by these challenges, we conducted a formative study with current business professionals, which revealed LLM's inadequate understanding of profession-level writing. The study also found a huge need to categorize a vast array of domain-specific conventions and requirements in writing tasks and make a structured taxonomy of such context. 

Therefore, we propose a novel approach of ``human-AI collaborative taxonomy construction'' for the use of domain-specific writing support. Such taxonomy construction can serve as a foundational step in developing AI writing assistants that can precisely interpret and generate text that aligns with the varied expectations of different professions. Also, based on the rigorous business expert validation procedures and a variety of human-AI interaction strategies, our approach can bolster the reliability of AI writing assistants customized to specific domains. For the rest of the paper, we interchangeably use the terms `domain-specific writing' and `profession-specific writing'.

%% file: Sections/Formative.tex
\section{Formative Study: Challenges of Using LLMs as Domain-specific Writing Assistants}

To assess the effectiveness of LLMs as domain-specific writing assistants, we conducted an initial formative study\footnote{This research has been approved by the Institutional Review Board (IRB) of the first author's institution.} with two professionals from the marketing and human resources (HR) sectors, each with over five years of field experience. These participants, referred to as P1 and P2, were asked to (1) provide inputs to GPT-4 to generate a writing template relevant to their field; (2) revise the template; and (3) categorize each edit according to predefined revision types outlined in a taxonomy \cite{du-etal-2022-understanding-iterative}, along with providing rationales behind their edits. The qualitative data was analyzed using thematic analysis and open coding techniques. For a detailed description of the study procedures, please refer to Appendix \ref{sec:formative-study}.

\begin{figure}[ht!]
    \centering
    \includegraphics[width=\linewidth, trim={1 0 0 0},clip]{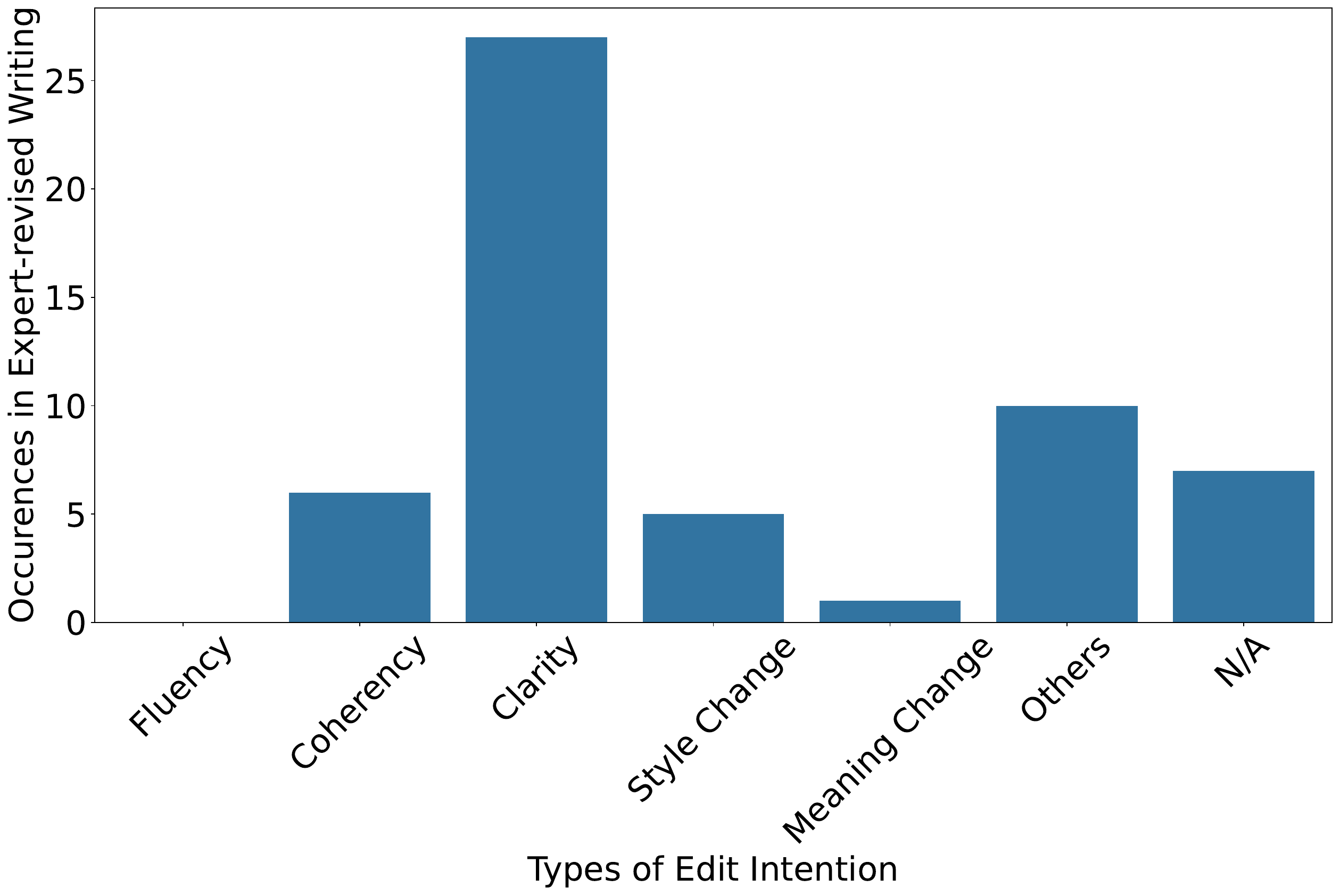}
    \caption{The distribution of revision intentions annotated by the study participants across six writing templates generated by GPT-4. }
    \label{fig:dist-edit-intention}
\end{figure}

\subsection{Limited Understanding of Nuances in Business Writing by LLMs}

We found that GPT-4's output often fails to align with the stylistic and linguistic expectations due to a lack of knowledge about domain-specific writing. First, the participants commonly criticized \textbf{clarity}, especially its verbosity (Figure \ref{fig:dist-edit-intention}): \textit{``The paragraph is too lengthy and I would like to get to the point right away.''} (P1) and  \textit{``... need to get driven to the CTA (Call-to-Action) as soon as possible... they don't want long emails..''} (P2). 
Second, the study participants highlighted (1) \textbf{lack of coherency} and (2) \textbf{style changes}. They commonly showed concern in GPT-4's lack of precision in word choice and organization structure that could \textbf{misinterpret or dilute the intended delivery in a business context}: \textit{``The word `essential' is subject to interpretation. Saying `absolutely necessary' might suggest the company is in such dire straits that people might leave the organization....''} (P1) and \textit{``Needs to call out sections to get the readers' attention. This is also the general style of a proposal.''} (P2). Third, all participants mentioned that GPT-4 created \textbf{its own narratives} that were not asked in the prompt and did not reflect objective details requested by the participant in the output. And finally, the participants who marked the ``Other'' category (Figure \ref{fig:dist-edit-intention}) pointed out LLMs' \textbf{lack of understanding about the objective of a writing task} (i.e., P2 mentioned \textit{``..a minute is to highlight what was discussed and action steps. Needs to be easy to read and call out sections a reader may care about.''}). 

From this formative study, we observed that current LLMs often lack nuanced comprehension of domain-specific expectations unless provided with detailed guidelines for their generated writing. Consequently, we identified the need to \textbf{develop a more sophisticated taxonomy of writing specific to various business domains}. This taxonomy will serve as a guideline, enhancing the pipeline for model training and enabling more tailored revision suggestions in domain-specific writing contexts.

%% file: Sections/New.tex
% \begin{figure*}[t!]
%     \begin{subfigure}{0.5\textwidth}
%     \centering
%         \includegraphics[width=1.0\linewidth]{images/step_1_newnew.pdf}
%         \caption{\textbf{STEP 1: Taxonomy Generation}}
%     \end{subfigure}
%     % \hfill
%     \begin{subfigure}{0.48\textwidth}
%     \centering
%         \includegraphics[width=1.0\linewidth]{images/step_2_newnew.pdf}
%         \caption{\textbf{STEP 2: Taxonomy Validation}}
%     \end{subfigure}
%     \vspace{2mm}
%     \begin{subfigure}{0.6\textwidth}
%     \centering
%         \includegraphics[width=1.0\linewidth]{images/step_3_newnew.pdf}
%         \caption{\textbf{STEP 3: Taxonomy Merging \& Testing}}
%         \vspace{1mm}
%     \end{subfigure}
%     \caption{An end-to-end pipeline of our three-step Human-AI collaborative taxonomy construction process. For each step, we portray several design implications for better human-AI interaction strategies that were described in Section 3.}
%     \label{fig:pipeline}
% \end{figure*}

\section{Human-AI Collaborative Taxonomy Construction}

In this section, we propose a novel task of ``human-AI collaborative taxonomy construction'' as an application to better develop a domain-specific AI writing assistant, following the standard procedures of taxonomy construction \cite{Nickerson2013AMF}. Grounded on the design implications of LLM-assisted qualitative analysis process \cite{Gao2023CoAIcoderET, Zhang2023RedefiningQA, Katz2023ExploringTE, Xiao2023SupportingQA, Jiang_2021, yan2023humanai}, we suggest the three-step approach with several design ideas (Figure \ref{fig:pipeline}).  Also, we present the promising results from our partial empirical testing of this three-step approach using a simplified scenario, as described in Appendix \ref{sec:initial-results}. 

% Grounded in the design implications of LLMs as an AI-assisted qualitative research tool \cite{Gao2023CoAIcoderET, Zhang2023RedefiningQA, Katz2023ExploringTE, Xiao2023SupportingQA, Jiang_2021, yan2023humanai}, we propose a novel task of ``human-AI collaborative taxonomy construction'' as an application to better develop domain-specific AI writing assistants. Here, we suggest the following three-step approaches with several design ideas (Figure \ref{fig:pipeline}).

% \begin{figure*}[t!]
%     \begin{subfigure}{0.49\textwidth}
%     \centering
%         \includegraphics[width=1.0\linewidth]{images/generation_1.pdf}
%         \caption{Average bias scores by model size}
%         \label{fig:scatter}
%     \end{subfigure}
%     % \hfill
%     \begin{subfigure}{0.49\textwidth}
%     \centering
%         \includegraphics[width=1.0\linewidth]{images/generation_2.pdf}
%         \caption{Breakdown of the proportional impact of each bias}
%         \label{fig:barplot}
%     \end{subfigure}
%     \caption{dd}
%     \label{fig:major_breakdown}
%     \vspace{-0.5cm}
% \end{figure*}

\subsection{STEP 1: Taxonomy Generation}
This process harnesses the capabilities of LLMs to generate all components of taxonomy from scratch\footnote{LLMs are trained on extensive and diverse datasets encompassing vast world knowledge \cite{liu2024understanding}, hence we task them with generating elements of the taxonomy.}, given the absence of domain-specific data or an existing taxonomy, which are typically required in traditional construction processes \cite{Nickerson2013AMF}. Based on the user inputs about a certain domain and sophistical designed output guidelines in the prompt design \cite{Mirowski2022CoWritingSA, wei2023chainofthought}, we \textbf{hierarchically generate each level of taxonomy}. Additionally, to address concerns of transparency and explainability in AI \cite{Dwivedi2023OpinionP, gomez2023designing, Gebreegziabher2023PaTATHC}, LLMs are prompted to \textbf{provide reasoning} for each generated element, increasing trustworthiness in the output.

\subsection{STEP 2: Taxonomy Validation}

To mitigate concerns about the dependence on artificially generated taxonomy as a beginning step of taxonomy construction, we perform multiple rounds of rigorous human expert validations and improvements, via \textbf{dialogue-based user interaction}. Inspired by several NLP works that proposed multiple LLM workarounds for a single task \cite{wang2023shepherd, Welleck2022GeneratingSB, Saunders2022SelfcritiquingMF}, our work suggests a novel approach of ``\textbf{LLMs-as-Mediators}'' as a human-AI interaction design during this validation process. Here, we employ multiple LLMs with different roles: (1) `Interviewer LLM' for asking questions and getting expert replies over multiple iterations through dialogue and (2) `Creator LLM' for generating the improved version of the taxonomy based on the received expert feedback from `Interviewer LLM'.

\subsection{Step 3: Taxonomy Merging \& Testing}

% To make a higher agreement among researchers, we simply ask LLMs to aggregate and combine the multiple finalized outputs by several experts working on the same domain. Also, LLMs will be asked to take into account that the elements of mixed output do not overlap with each other but look comprehensive. Utilizing LLMs will reduce individual bias and increase the reliability of the output. Then, taking the current practices of the taxonomy evaluation process \cite{Nickerson2013AMF, shah2024using}, we ask both human experts and LLMs to annotate writing templates with the constructed taxonomy, separately. If the inter-coder reliability (ICR) between human coders is high and the ICR between GPT-4 and expert annotations is also high, then we confirm the substantial reliability of the generated taxonomy for the specific domain. 

To foster greater consensus among researchers, we propose a method where LLMs aggregate and merge multiple finalized outputs from several experts within the same domain. Moreover, LLMs will ensure that the elements of the combined output are comprehensive and non-overlapping. By leveraging LLMs, we aim to mitigate individual biases and enhance the reliability of the generated taxonomy. Subsequently, following established practices in taxonomy evaluation processes \cite{Nickerson2013AMF, shah2024using}, both human experts and LLMs will independently annotate writing templates using the constructed taxonomy. We will assess the inter-coder reliability (ICR) between human coders and between LLM and expert annotations. High ICR scores between both sets of annotators would confirm the substantial reliability of the taxonomy for the specific domain.

%% file: Sections/Results_Conclusion.tex
\section{Future Plans}

% With the proposed approach of human-AI collaborative taxonomy construction, we plan to implement the following work as our future work. First, we will develop an interface with the three steps including user input (STEP 1) and a chat component (STEP 2) for better human-AI interaction. In addition, we will begin larger-scale expert recruitment via a freelancing platform (e.g., Upwork) to conduct the three steps and validate the effectiveness of such human-AI collaboration. After we collect amounts of taxonomies per domain, we will begin the model training process with open-source and closed LLMs to develop domain-specific writing assistants. With all these plans, we expect to see an improvement in the quality of AI wrii

% \minhwa{will mention a brief future plan in 2-3 sentences. (1) interface development; (2) more advanced prompting technique; and (3) expert recruitment and experiment}

Our plan involves developing a user-friendly web application with two main components: (1) a hierarchical taxonomy display and (2) a chat dialogue interface with an Interviewer LLM. Through iterative dialogue, users will validate and refine the taxonomy, while recent prompting techniques like ReAct \cite{yao2023react} and ART \cite{paranjape2023art} will define desired LLM behaviors. Furthermore, we plan to conduct the experiments with open-source models (e.g., Mistral \cite{jiang2023mistral}, LLaMA \cite{touvron2023llama}, OLMo \cite{groeneveld2024olmo}, etc.) in addition to GPT-4. Lastly, we will recruit domain experts via a freelancing platform (e.g., Upwork\footnote{\href{https://www.upwork.com/}{https://www.upwork.com/}}) for multiple human studies to ensure the effectiveness of our approach. 

%By synergizing human expertise with LLM capabilities, our goal is to pioneer domain-specific writing support tailored to various professional sectors. Ultimately, this approach aims to enhance writing assistance by addressing specific stakeholder needs.


% \begin{acks}
% Generous support by Accenture Lab
% \end{acks}

%% file: Sections/Appendix.tex
\appendix

\section{More About Formative Study}\label{sec:formative-study}

\subsection{Recruitment Process}

We recruited 2 industry professionals (referred to as P1 and P2) via the U.S.-based freelancing platform `Upwork.' Note that the study was approved by the Institutional Review Board of the first author's institution. When recruiting the participants, we set the eligibility of participation as follows: (1) at least 3 years of working experience in their business sector; (2) fluent in English (Native Speaker Preferred); and (3) prior experience in several business-specific writing tasks such as email, proposals, and meeting minutes. In addition to meeting the three requirements, we also asked the applicants to answer the screening question: \textit{``Please describe your level of experience in your work group, including specific project details.''}. Based on the manual review of the responses, we invited the two participants (P1 and P2) who met all the requirements of our study. After the successful completion of the study, the participants received a compensation of 20 USD for their participation in a 1-hour duration of our formative study. Table \ref{table:participant} shows the detailed background of the study participants we recruited. 

\begin{table*}[ht!]
\centering
\resizebox{\textwidth}{!}{
\begin{tabular}{@{}ccccccc@{}}
\toprule
                  & \textsc{Gender} & \textsc{Residence} & \textsc{Working Domain} & \textsc{Years of Experience} & \textsc{English Proficiency} & \textsc{Disability Status}\\ \midrule
P1   & Male & United States & Human Resources (HR) & 14 &  Native & No    \\
P2   & Female & United States & Marketing & 8 & Native & No \\ \bottomrule
\end{tabular}
}
\vspace{1em}
\caption{Detailed background information of the formative study participants.}
\label{table:participant}
\end{table*}

\subsection{Experiment Procedures} \label{sec:formative-procedure}

The formative study involved three steps. First, the participant provided their background information: (1) last name (for data file identifier purposes only); (2) domain group; (3) the type of writing task (e.g., meeting minutes, emails, and project proposals); and (4) the content of the writing in fully complete English sentences. Based on these user inputs, the GPT-4 integrated into the interface generated a writing template. When the generated writing was present, the participants were asked to revise the writing and to click the button to store all the revisions securely. 

Next, the participants were prompted to move to the next page of the interface, where their revisions were visualized with the additions in red and the deletions in strikethrough. This page was prepared to let the participants check whether their revisions were properly stored and visualized; if not, we asked them to go back to the previous step and provide revisions again. 

Lastly, after they confirmed that their revisions were successfully stored and visualized, we let the participants move to the last page of the interface. Here, they were given a list of instructions on how to annotate their sentence-level edits with the provided types of revision intentions that were established by a previous work \cite{du-etal-2022-understanding-iterative}. After they completed the annotation of every edit they made in the GPT-4 generated writing, they were asked to leave the interface by closing the tab of the interface in their local search engine (e.g., Chrome, Safari). Our interface was developed to store all information produced during the study into our password-protected server - (1) revision history with the annotations; (2) the original GPT-4 writing; and (3) the user inputs. Please note that we used the last name information only for data identifier purposes (i.e., renaming the data file with the last name) and did not collect or use any personally identifiable information during the study. 

\subsection{Interface Development}
We used Streamlit and Python to develop a web interface for the formative study process. Then, we made the web interface publicly accessible by hosting it on a website using ngrok. This enabled the participants to access the interface using their local search engine. The code repository is available at the following GitHub repository: \url{https://github.com/minnesotanlp/co-taxonomy}. Figure \ref{fig:interface-1} is a screenshot of the interface shown to the participants for the first step described in Appendix \ref{sec:formative-procedure}.

\begin{figure*}[ht!]
    \centering
    \includegraphics[width=\textwidth]{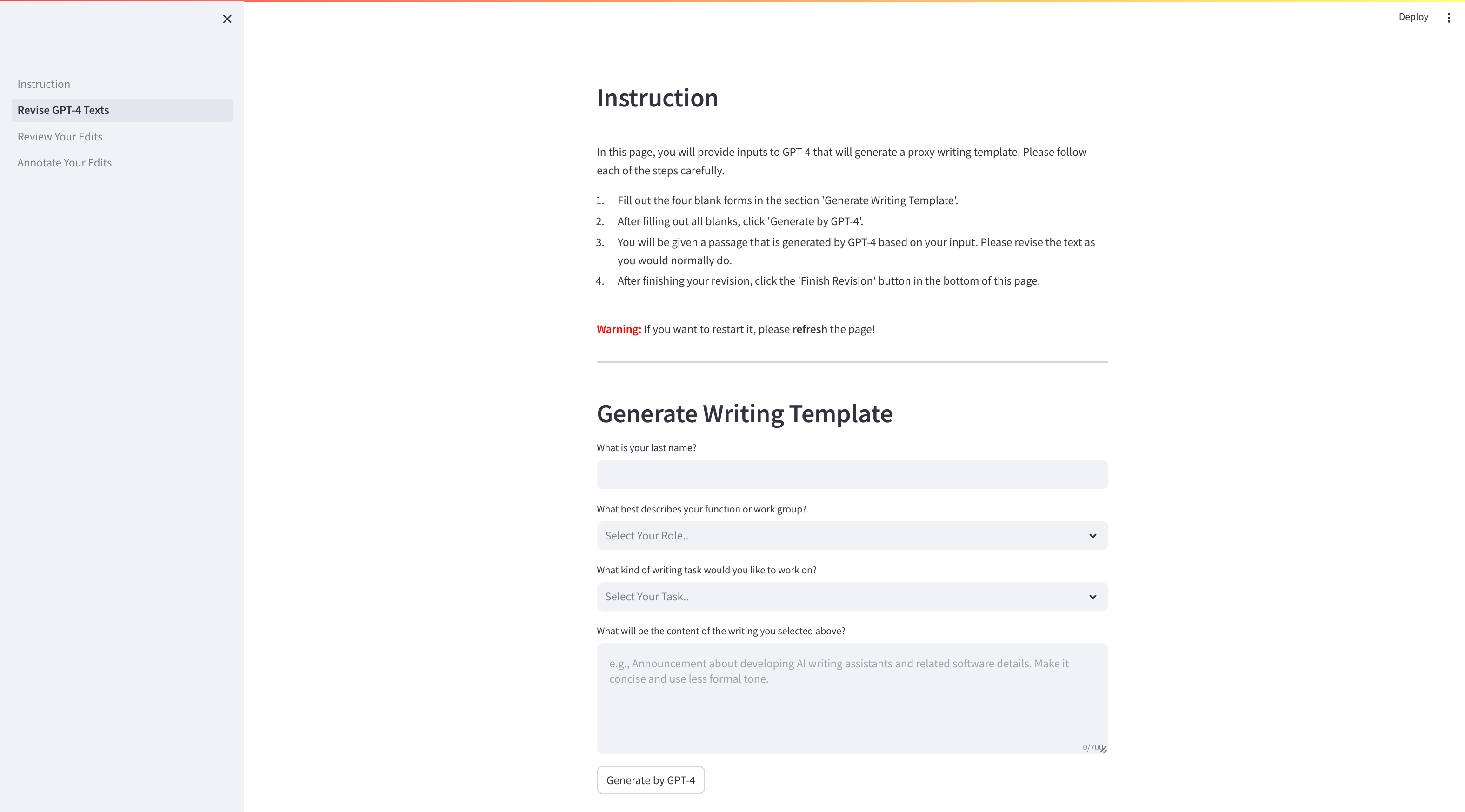}
    \caption{The interface for the first step, where a participant provided their background information that GPT-4 then used to generate a writing template.}
    \label{fig:interface-1}
\end{figure*}
% \begin{figure*}[ht!]
%     \centering
%     \includegraphics[width=\textwidth]{images/Review Your Edits.pdf}
%     \caption{The interface for the second step, which visualizes the participant's revisions in the GPT-4 generated writing.}
%     \label{fig:interface-2}
% \end{figure*}
% \begin{figure*}[ht!]
%     \centering
%     \includegraphics[width=\textwidth]{images/Annotate Your Edits.pdf}
%     \caption{The interface for the last step, where a participant annotates each edit according to predefined revision types \cite{du-etal-2022-understanding-iterative}, in addition to providing rationales behind their edits.}
%     \label{fig:interface-3}
% \end{figure*}

\section{Initial Results} \label{sec:initial-results}

We have conducted empirical testing of our human-AI collaborative taxonomy construction process, partially executing STEP 1 and 2 using a simplified scenario focused on developing a taxonomy of text revision intentions, specifically for email writing within the legal domain. 
% For a comprehensive description of the process, please refer to the Appendix \ref{sec:initial-results}.

\begin{table*}[h]
\resizebox{\textwidth}{!}{%
\begin{tabular}{l|l|l}
\toprule
\textbf{Intention} & \textbf{Description} & \textbf{Examples}\\
\midrule
\Large \multirow{8}{*}{\begin{tabular}[c]{@{}l@{}}\textit{Legal Argument} \\ \textit{Strengthening}\end{tabular}} & \multirow{2}{*}{\Large \begin{tabular}[c]{@{}l@{}}Adding supporting legal \\ precedents to reinforce an \\ argument.\end{tabular}}   & \begin{tabular}[c]{@{}l@{}}\large The case we are handling has similarities with other cases. \\ $\rightarrow$ \large The case we are handling is similar to Smith v. Jones, \\ \large where the court held a comparable view on contractual obligations.\end{tabular} \\ \cline{3-3} 
 &    & \multicolumn{1}{l}{\large \begin{tabular}[c]{@{}l@{}}Our argument in this lawsuit is strong. \\ $\rightarrow$ Our argument, bolstered by the precedent set in Brown v. Board of Education, \\is particularly strong in advocating for equal rights. \end{tabular}} \\ \cline{2-3} 
 & \multirow{2}{*}{\Large \begin{tabular}[c]{@{}l@{}}Integrating additional factual\\ evidence to solidify a legal stance.\end{tabular}} &  {\large \begin{tabular}[c]{@{}l@{}} Our client's position in this matter is legally sound. \\ $\rightarrow$ Our client's position is legally sound, supported by the financial records \\ and witness statements provided. \end{tabular}} \\ \cline{3-3} 
 &    & \multicolumn{1}{l}{\large \begin{tabular}[c]{@{}l@{}}This case is straightforward. \\ $\rightarrow$ This case is straightforward, as evidenced by the detailed timeline \\ of events and corroborating emails.\end{tabular}} \\ \cline{2-3} 
& \multirow{2}{*}{\Large \begin{tabular}[c]{@{}l@{}}Enhancing the persuasiveness of \\ the argument by refining logical \\ reasoning.\end{tabular}}   &   {\large \begin{tabular}[c]{@{}l@{}}We believe our client is not liable. \\ $\rightarrow$ Our client is not liable, as logically, the responsibility falls \\ on the contractor, given the terms of the agreement.\end{tabular}} \\ \cline{3-3} 
 &  & \multicolumn{1}{l}{\large \begin{tabular}[c]{@{}l@{}}This case should be dismissed. \\ $\rightarrow$  This case should be dismissed, considering the lack of \\ causation between our client's actions and the alleged damages\end{tabular}} \\ \cline{2-3} 
 & \multirow{2}{*}{\Large \begin{tabular}[c]{@{}l@{}}Incorporating expert testimony \\ to bolster legal claims.\end{tabular}} &   {\large \begin{tabular}[c]{@{}l@{}}Our stance on the patent infringement is valid. \\ $\rightarrow$ Our stance is strengthened by the expert testimony of Dr. Smith, \\ a renowned patent specialist.\end{tabular}}\\ \cline{3-3} 
  &   & \multicolumn{1}{l}{ \large \begin{tabular}[c]{@{}l@{}}The damages claimed are excessive. \\ $\rightarrow$ The damages claimed are excessive, as per the assessment of \\ leading industry expert John Doe.\end{tabular}} \\
\bottomrule
\end{tabular}%
}
\vspace{1em}
\caption{An example of the GPT-4 generated taxonomy of text revision for legal domain, through STEP 1. The proposed revisions by the model are indicated by the arrows.}
\label{table:step1-example}
\end{table*}

% We have empirically tested our design for the human-AI collaborative taxonomy construction process by partially running STEP 1 and 2, based on a toy scenario of developing a taxonomy of writing intentions (e.g., revisions) for email writing in the legal domain. See Appendix for the full descriptions of the process. 

% Table \ref{table:step1-example} presents one example of the generated elements of all levels of the taxonomy through STEP 1, where the descriptions and the corresponding examples are relevant to the legal domain and easy to understand. Through STEP 2, the `Interviewer LLM' asked one of the authors with the following question: ``What do you think can be added as an additional \textit{description} category if not comprehensive?''. Based on the author's reply that they need to include some rebuttal provision to legal arguments, the Interviewer LLM suggested another description of ``Addressing Counterarguments'' with its reasoning saying that ``Proactively identifying potential weaknesses or counterarguments to the legal position''. While the results are not expert-validated yet, we believe that this line of results can support the effectiveness of our approach and that we will run the entire process with larger-scale experiments and more advanced multi-step prompting techniques such as ReACT \cite{yao2023react} or ART \cite{paranjape2023art}. 

Table \ref{table:step1-example} showcases an example of the generated taxonomy elements across all levels, featuring descriptions and examples pertinent to the legal domain, ensuring clarity and relevance. During STEP 2, the `Interviewer LLM' engaged one of the authors, posing the question: ``What do you think can be added as an additional \textit{description} category if not comprehensive?''. Upon the author's suggestion to include provisions for addressing counterarguments in legal arguments, the Interviewer LLM proposed the addition of a ``Addressing Counterarguments'' category, along with a rationale emphasizing proactive identification of potential weaknesses or counterarguments to legal positions. Although these results are pending expert validation, they signify the promise of our approach. We plan to further validate this methodology through larger-scale experiments and incorporate advanced multi-step prompting techniques such as \textsc{ReACT} \cite{yao2023react} or \textsc{ART} \cite{paranjape2023art}.

\subsection{Hierarchical Prompt Design (STEP 1)}

We present the designed prompts for GPT-4 to generate all levels of taxonomy (according to STEP 1), using a hierarchical prompt chaining technique inspired by \citet{wei2023chainofthought, Mirowski2022CoWritingSA}. Listing \ref{lst:label-prompt} present the prompt for the `Intention' category generation.
% ; Listing \ref{lst:description-prompt} for `Description' category generation; and Listings \ref{lst:example-prompt-1} and \ref{lst:example-prompt-2} for `Example' generation. 

\subsection{Questions for Interviewer LLM (STEP 2)}

For the dialogue-based validation during STEP 2, we brainstormed several questions for `Interviewer LLM' to check the following aspects of the `Description' of the taxonomy. These aspects were partially drawn from \citet{shah2024using} that suggested some evaluation criteria for the LLM-generated taxonomy of user intents in the information retrieval (IR) domain. 

\begin{itemize}
    \item Consistency: ``Do all descriptions for label not overlap in the meaning? If not, which description seems ambiguous/duplicate?''
    \item Clarity: ``Do the descriptions provide clear understanding? If not, which one is unclear and how should it be improved?''
    \item Practicality: ``Is there any description that is not typically considered in your domain? Which one?''
    \item Comprehensiveness: ``What do you think can be added as an additional \textit{description} category if not comprehensive?''
\end{itemize}

\begin{listing*}[!ht]
\begin{minted}[fontsize=\footnotesize, frame=single, breaklines]{markdown}
f'''
You're a helpful assistant helping a user generate the labels of edit intentions in their writing task of {email}. 
The user is a business professional in the following domain: {legal domain}. 

### Output Guidelines: 

- Suppose that the user revises their original text of {email} for better communications within their domain group (e.g., a marketing person for other team members in the marketing team; an HR worker writing for the HR team, etc.)
- Please consider intentions of revising contents, specifically for the domain of {legal}.
- In addition to domain-specific content-related taxonomy, include general syntactic revision intentions, such as grammar corrections, style coherency, clarity, etc. 
- Please generate labels of feasible scenarios as much as you can. It should be more than 10 labels at least. 
- For each generated intention, prepend the token <label> and append </label>.
- Please consider that all those generated labels must be mutually exclusive (e.g., no overlapping between labels). 
- Please consider that all those generated labels must be collectively exhaustive (e.g., cover every possible case of revision intentions in practice). 
- Once you finish generating all possible labels of revision intention of {email} for the domain of {legal}, end your response with the tag <end>.

### Example 

Given the domain of marketing and the writing task of email, the example looks as follows: 

```
<label> Target Engagement </label>
<label> Visual Content Integration </label> 
<label> Data-Driven Insights Presentation </label>
....
<end>
```
### Output
'''
\end{minted}
\caption{A prompt for GPT-4 to generate the `Intention' category. }
\label{lst:label-prompt}
\end{listing*}